\documentclass[manuscript]{aastex}
\shorttitle{Infrared studies of X Per}
\shortauthors{Mathew et al.}

\begin{document}

\title{Studies of the Be star X Persei during a bright infrared phase}

\author{Blesson Mathew, D. P. K. Banerjee, Sachindra Naik and N. M. Ashok}
\affil{Astronomy and Astrophysics Division, Physical Research Laboratory,
Navrangapura, Ahmedabad - 380 009, Gujarat, India}
\email{blessonmathew@gmail.com}

\begin{abstract}
We present multi-epoch photometric and spectroscopic near-infrared
observations of the Be star X Persei in the $JHK$ bands covering the wavelength range
1.08 to 2.35 $\mu$m. Combining results from our earlier studies with the
present observations, it is shown that the equivalent widths and line fluxes
of the prominent H{\sc i} and He{\sc i} lines anti-correlate with the strength of
the adjacent continuum. That is, during the span of the observations,
the $JHK$ broad band fluxes increase while the emission equivalent width
values of the H{\sc i} and He{\sc i} lines decrease (the lines under consideration
being the Paschen and Brackett lines of hydrogen and the 1.7002 $\mu$m
and 2.058 $\mu$m lines of helium). Such an  anti-correlation effect is not
commonly observed in Be stars in the optical; in the infrared this may
possibly be the first reported case of such behavior. 
We examine different mechanisms that could possibly cause
it and suggest that it could originate due to a  radiatively warped,
precessing circumstellar disk. It is also shown that during the course of
our studies X Per evolved to an unprecedented state of high near-IR
brightness with $J, H, K$ magnitudes of 5.20, 5.05 and 4.84 respectively.
\end{abstract}


\section{Introduction}

X Persei/HD 24534 is a bright and variable Be star, the
optical/near-IR counterpart of the low-luminosity X-ray source 4U 0352+30,
which belongs to the class of Be/X-ray binaries
\citep{br72,roc97}. \citet{nor91} identified a disk loss state in X Per during 1989-1990
while \citet{roc93} presented  evidence for a similar `extended low-luminosity state'
during 1974-1977. \citet{fab92} analyzed the data during the disk-less phase
and estimated the spectral type to be O9.5 {\sc iii}e and set an age of 6 Myr
for the X Per system. The distance and rotation velocity ($v~sin~i$) are
estimated to be 1300 pc and 200 km s$^{-1}$ respectively \citep{fab92}.
The orbital period of the system was inferred from $RXTE$ data to be 250 days
while the eccentricity is 0.11
and orbital inclination angle is in between 26 and 33 degrees \citep{del01}.

\citet{mat12b} did infrared photometric and spectroscopic studies of X
Per during 2010 December 4 to 2011 February 8 and
found the object to be in a photometrically bright state in the
near-IR compared to earlier recorded values.
Large optical depth effects were also observed in the H{\sc i} recombination lines
as implied from the deviation of Paschen and
Brackett line strengths from recombination Case B values. The object was found to show
infrared excess which was modeled using a free-free excess from
the disk to allow estimating a mean value of electron density, $n_e$ = 4 $\times$
10$^{11}$ cm$^{-3}$, which is consistent with the values estimated in Be stars.

Because of the  steady brightening trend in the $JHK$ bands,
we continued monitoring X Per from Mt. Abu Infrared Observatory which
has led to two interesting results that are reported here.
First, the object is found to have evolved to an unprecedented state of high
near-IR brightness. This  brightening is worth reporting as it should
interest other observers and encourage further monitoring of the object.
The second significant result is the observation of an
anti-correlation between the line equivalent widths (and also line fluxes)
with the continuum brightness as measured by the IR magnitudes.

A search in the literature indicates that this type of inverse 
correlation is not too frequently  seen in Be stars. \citet{sig13} 
have made a detailed study of the dependence of H$\alpha$ equivalent widths 
on the V band magnitudes during long term variations of classical Be stars. 
They point out the examples of a few stars which have exhibited such an
inverse correlation  namely: 4 Her (B9e; \citealt{Koubsky97}), 88 Her (B7pshe; 
\citealt{Doazan82}), V1294 Aql (B0Ve; \citealt{Horn82}) and Eta Cen (B1.5IVne; \citealt{Stefl95}).

Another  notable example is that of the
Be star \object{delta Sco} in which \citet{mir03} identified an
anti-correlation between the H$\alpha$ line flux and $V$-band brightness of
the star after its 2001 periastron passage, during which it transitioned from a
normal B-type star to a Be star.
This increase in H$\alpha$ line flux with a fading in $V$-band flux
was explained in terms of episodes of enhanced mass loss from the star.
However, this explanation was not found entirely satisfactory by
\citet{car06} for several reasons. They discussed and studied
several alternative causes for the observed anti-correlation behavior
but concluded by admitting that a convincing explanation could not be
offered by their model (see section 4 in Carciofi et al. 2006).
They suggested that rather than a static model of the Be star circumstellar disk,
a complex dynamic model with a significant change in mass-loss rate and circumstellar
geometry is needed to explain this phenomenon. \citet{mar01} identified an anti-correlation between the H$\alpha$
equivalent width and $V$-band magnitude of the luminous blue variable
(LBV) star P Cygni. \citet{pol12} recently verified this anti-correlation as part of an
international observing campaign where they did contemporaneous $V$-band
photometry and H$\alpha$ equivalent width measurements. In view of the limited examples
of this phenomenon seen in Be stars, it becomes necessary to report
and if possible understand the observed behavior of X per.

\citet{liu01} observed X Per during the bright phase in 1994-1995 and found
a rise in infrared flux values with a peak in 1994 September 19 followed by
the fading of light curve. The emission strength of H$\alpha$ also increased
during this time but reached the peak only after the fading in infrared light
curve started. They developed an expanding ring model to explain this
phenomenon in which they proposed the infrared excess and H$\alpha$ line
emission to arise from different regions of the envelope and their maximum
values are reached at different evolution stage of the envelope.
They also found that most of the infrared
continuum emission came from within a region of 2 stellar radii while
the maximum H$\alpha$ emission arose from 6 stellar radii.

The present work presents $JHK$ photometry that show the progress of X Per to
its present high level of brightness. Multi-epoch $JHK$ spectra are also presented from
which we analyze the temporal evolution and dependence of the line equivalent widths and line fluxes on the near-IR magnitudes.

\section{Observations \& Analysis}

The infrared photometric and spectroscopic observations of X Per were carried out from
the 1.2-m Mt. Abu telescope, operated by the Physical Research Laboratory.
Photometry in the $JHK$ bands was done in photometric sky conditions using
the imaging mode of the Near-Infrared Imager/Spectrometer which employs a
256 $\times$ 256 HgCdTe NICMOS3 array. Several frames, in five
dithered positions offset typically by 20 arcsec, were obtained for
both the program object and a selected standard star
(\object{SAO 56762}; spectral type A5V) in each of the $J, H, K$
filters. The procedure for the near-IR photometric observations and the subsequent
reduction and analysis of data followed a standard procedure described for example
in \citet{ban02}. The log of the photometric
observations and the $JHK$ magnitudes with the estimated errors are shown in
Table 1.

The near-IR spectra were obtained at a resolution of $\sim$1000 in each of the
$J, H, K$ bands using the NICMOS3 array. The
spectra were extracted using IRAF and the wavelength calibration was
done using a combination of OH sky lines and telluric lines that
register with the stellar spectra. We have also observed the
standard star \object{SAO 56762} at a similar airmass to the object.
The data were reduced using  standard IRAF tasks and the details
of the reduction procedure is given in the earlier study of X Per by \citet{mat12b}.
The log of the spectroscopic observations is given in Table 2.

%table 1
\begin{table}
\begin{center}
\caption{Journal of the photometric observations of X Per\label{tbl-1}}
\begin{tabular}{@{}lccccc@{}}
\tableline\tableline
 Date & MJD & Exp.time & \multicolumn{3}{c}{Photometry}\\
      &     & $(J, H, K)$ & \multicolumn{3}{c}{Magnitudes}\\
      &     & (s)  & $J$ & $H$ & $K$ \\
\tableline
2011 &  &   &  & & \\
Oct. 14 & 55848.8 &  (0.25,0.2,0.3) & 5.30$\pm$0.01 & 5.22$\pm$0.02 & 4.96$\pm$0.01 \\
Oct. 28 & 55862.8 &  (0.5,0.5,0.5) & 5.37$\pm$0.03 & 5.18$\pm$0.02 & 4.98$\pm$0.02 \\
Oct. 29 & 55863.8 &  (0.3,0.3,0.3) & 5.30$\pm$0.02 & 5.22$\pm$0.01 & 5.01$\pm$0.03 \\
Oct. 30 & 55864.9 &  (0.5,0.4,0.4) &     \nodata   & 5.27$\pm$0.03 &  \nodata      \\
Oct. 31 & 55865.8 &  (0.2,0.2,0.2) & 5.35$\pm$0.03 & 5.20$\pm$0.02 & 5.01$\pm$0.03 \\
Nov. 01 & 55866.8 &  (0.2,0.2,0.2) & 5.39$\pm$0.02 &  \nodata      & 5.05$\pm$0.06 \\
Nov. 06 & 55871.8 &  (0.3,0.3,0.3) & 5.35$\pm$0.02 & 5.20$\pm$0.02 & 4.99$\pm$0.02  \\
Nov. 07 & 55872.9 &  (0.3,0.3,0.3) & 5.34$\pm$0.02 & 5.19$\pm$0.04 & 4.95$\pm$0.02 \\
Nov. 10 & 55875.9 &  (0.3,0.3,0.3) & 5.34$\pm$0.04 & 5.19$\pm$0.03 & 4.96$\pm$0.03 \\
Nov. 11 & 55876.9 & (0.3,0.3,0.3) &  5.35$\pm$0.02 & 5.18$\pm$0.01 & 4.99$\pm$0.03 \\
Nov. 12 & 55877.9 & (0.25,0.25,0.25) & 5.36$\pm$0.01 & 5.20$\pm$0.03 & 5.02$\pm$0.04 \\
Nov. 21 & 55886.9 & (0.25,0.25,0.25) & 5.28$\pm$0.05 & 5.26$\pm$0.01 & 4.96$\pm$0.06 \\
Dec. 01 & 55896.9 & (0.3,0.3,0.3) & 5.34$\pm$0.02 & 5.24$\pm$0.03 & 4.99$\pm$0.03 \\
Dec. 13 & 55909.0 & (0.25,0.25,0.25) & 5.43$\pm$0.04 & 5.35$\pm$0.01 & 4.94$\pm$0.02 \\
2012 &  &   &  & & \\
Jan. 07 & 55934.0 & (0.25,0.25,0.25) & 5.28$\pm$0.03 & 5.19$\pm$0.02 & 4.96$\pm$0.02 \\
Feb. 28 & 55985.8 & (0.25,0.25,0.25) & 5.32$\pm$0.03 & 5.17$\pm$0.02 & 4.94$\pm$0.02 \\
Mar. 08 & 55994.5 & (0.25,0.25,0.25) & 5.27$\pm$0.03 & 5.15$\pm$0.03 & 4.92$\pm$0.03 \\
Mar. 27 & 56013.8 & (0.5,0.5,0.5) & 5.28$\pm$0.02 & 5.12$\pm$0.01 & 4.90$\pm$0.02 \\
Mar. 28 & 56014.8 & (0.5,0.5,0.5) & 5.24$\pm$0.04 & 5.10$\pm$0.02 & 4.93$\pm$0.04 \\
Apr. 06 & 56023.8 & (0.3,0.3,0.3) & 5.21$\pm$0.02 & 5.05$\pm$0.04 & 4.84$\pm$0.02 \\
\tableline
\end{tabular}
\end{center}
\end{table}

%figure 1
\begin{figure}
\plotone{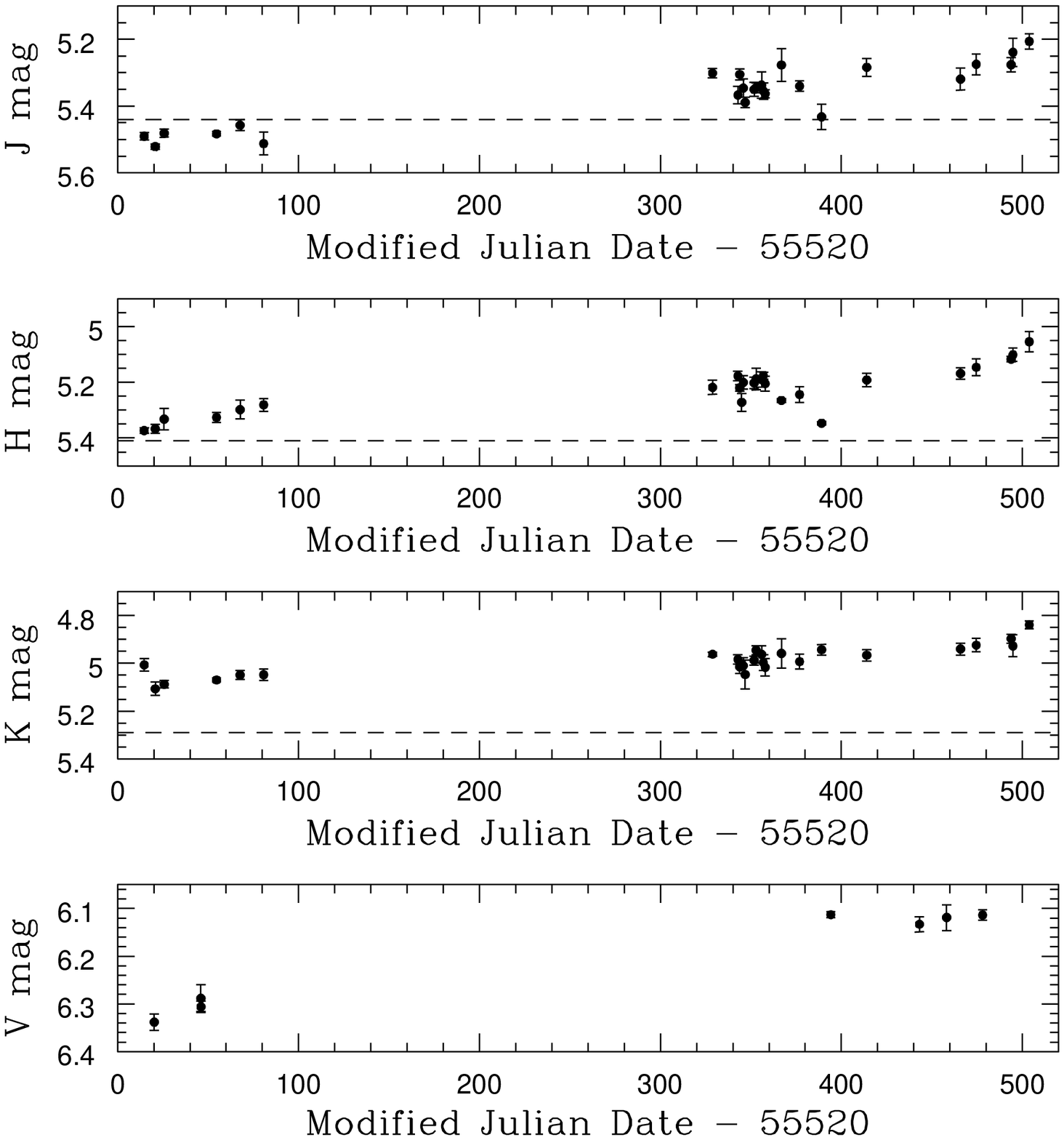}
\caption{Near-IR and optical light-curve of X Per during the period 2010 December 04 to
2012 April 05. Our set of $JHK$ magnitudes are shown as filled circles with
the error estimates. The $J, H, K$ values 5.44, 5.41 and 5.29 from
\citet{tel98}, which are the earlier known  brightest IR magnitudes
recorded for the object, are shown in dashed lines.
The $V$-magnitudes are taken from AAVSO database.\label{fig1}}
\end{figure}

\section{Results}

\subsection{Photometry}

The photometric observations of X Per were conducted during the period 2011 October 14
to 2012 April 6. The log of the observations along with the
estimated $JHK$ magnitudes are given in Table 1. For the construction of light
curve, we have included the $JHK$ values obtained during 2010 December 4 -- 2011
February 8; that were reported earlier in \citet{mat12b}.
The $JHK$ light curve after including the whole set of observations
(26 epochs) spanning a period of 489 days is shown in Figure 1.
It can be seen that the latest $JHK$ magnitudes are 5.21, 5.05 and 4.84,
obtained on 2012 April 6.
The 2MASS $JHK$ magnitudes of the object are 6.149, 6.073, 5.920 and hence
the present estimates are about 1 magnitude brighter than these values.
\citet{tel98} have compiled the most comprehensive set of
$JHK$ magnitudes of X Per for a period of 25
years. The brightest $JHK$ magnitudes reported by them are 5.44, 5.41 and 5.29
respectively, which was recorded  during 1994 September 16 -- 20, and is
shown as a dashed line in Figure 1.
\citet{liu01} also reported a bright state in X Per with $JHK$ magnitudes of 5.32,
5.26, 5.19 on 1994 September 19.
The $JHK$ values presented in this paper are 0.2 -- 0.3 mag brighter than these
values and hence show  that X Per is at an all-time bright state in the near-IR.

In order to see whether a contemporaneous brightening event occurs in the optical also,
we have looked for the $V$-band CCD data of X Per in the AAVSO database.
We could locate only 7 sets of photoelectric/CCD observations during the
period of our IR observations; though several visual estimates were available
these were not considered due to the lack of accuracy.
The light curve is plotted and is shown in the bottom panel of Figure 1.
The $V$-band light curve also brightened up and the magnitude changed from
6.338, when observed on 2010 December 10, to $V$ = 6.114 on 2012 March 12.
It can be seen that the magnitude change in $V$-band is 0.22, roughly similar
to those in the $J$, $H$, $K$ bands which are 0.29, 0.32, 0.25 respectively.
The magnitude excess in all the bands are similar even though we expected a
lesser value in $V$ if the excess emission is purely from the circumstellar
disk.

%figure 2
\begin{figure}[h]
\includegraphics[scale=0.65,angle=-90]{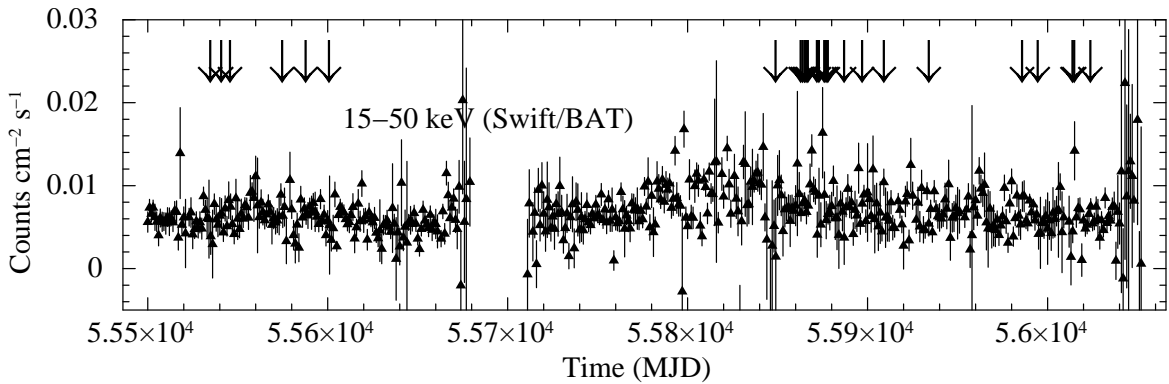}
\caption{The $Swift/BAT$ X-ray light curve of X Per in the 15 -- 50 keV energy band
covering the period of our infrared observations. The epochs corresponding to
infrared photometric observations are indicated by arrows.\label{fig2}}
\end{figure}

X Per belongs to the class of persistent X-ray binaries
where the neutron star does not pass through the
circumstellar disk during its periastron passage due to the low value of
eccentricity. Hence one does not expect normal Type I outburst in these systems.
But there are reports of feeble X-ray outbursts in X Per by \citet{lut12}
from a continuous monitoring during 2001-2011 using the data from $RXTE$,
$Swift$ and $INTEGRAL$ observatories.
Using simultaneous $V$-band magnitudes from AAVSO database
along with the X-ray data, they did not find any
significant correlation between the X-ray and optical light curve behavior.
The $Swift/BAT$ X-ray light curve of the system in 15 -- 50 keV
energy band is shown in Figure 2.
The epochs corresponding to our infrared photometric observations are marked in
the figure. No detectable flare in X-rays is visible during the
period of infrared observations i.e., from 2010 December 4 to 2012 April 6.
The orbital period of the system is 250 days \citep{del01} and hence the
present set of observations spanning 489 days cover almost two orbits.
Hence it can be concluded that the  neutron star does not influence or
have any significant role in the observed behavior reported here.
In the case of X Per, no disk loss event was reported after 1989.
If the present brightening is due to the deposition of material in the disk,
then it could lead to the expansion of the disk which may
result in a future X-ray outburst event. However, at present the possibility of such
an outburst in X Per appears to be low \citep{rei99}.

\subsection{Spectroscopy}

In the present study we obtained  $JHK$ spectra for 27 nights between 2011 October 14 to
2012 April 5 (from here on referred to as epoch 2).
The log of the observations for these data is shown in Table 2.
To prevent over-crowding in the display of these spectra,
we present in Figures 3, 4 and 5 the $J$, $H$ and $K$ spectra
respectively for 10 selected dates which are reasonably
equi-spaced over the  observation span of epoch 2 and give a
good idea of the spectral evolution. The $JHK$ spectra of X Per
are dominated by the recombination emission lines of hydrogen
and helium. The prominent recombination lines present in the spectra are
Pa$\beta$ 1.2818 $\mu$m, Pa$\gamma$ 1.0938
$\mu$m lines in Paschen series, Br$\gamma$ 2.1656 $\mu$m
and Brackett 10 -- 18 lines in Brackett series.
Of the helium lines He{\sc i} 1.0830 $\mu$m, 1.7002 $\mu$m and 2.058 $\mu$m are
prominent which are present in $J$, $H$ and $K$ band spectra respectively.
The O{\sc i} 1.1287 $\mu$m line is also seen whose strength is known to
be influenced by the Lyman $\beta$ fluorescence process (e.g., Mathew et al. 2012a). The
equivalent widths of the prominent lines are given in Table 2 -- the typical
error in the measurement of the equivalent width values is 10 \%.

%figure 3
\begin{figure}
\plotone{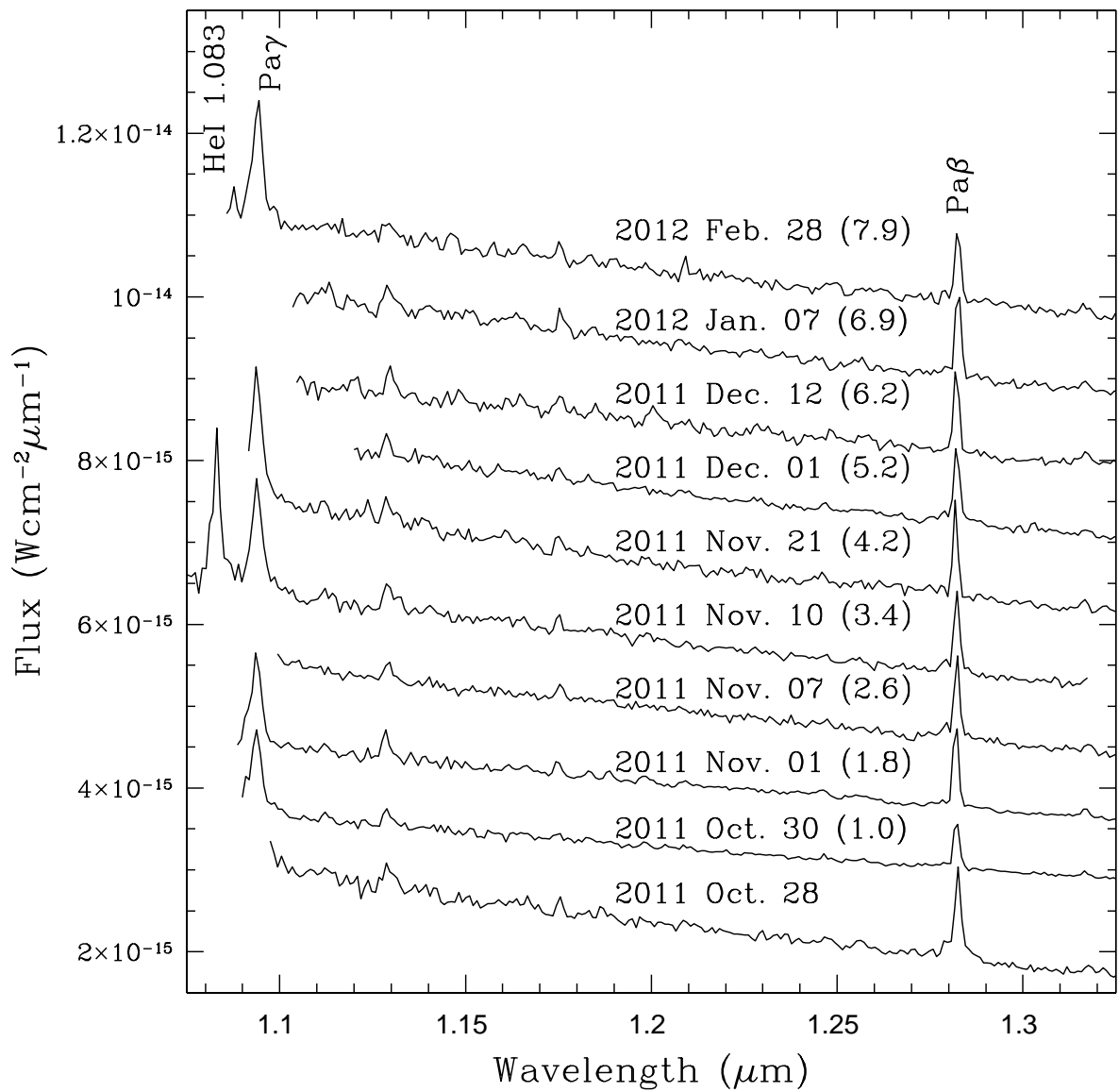}
\caption{Flux calibrated $J$ band spectra of X Per are displayed at different
  epochs with an offset between adjacent spectra for clarity. The amount of offset in
  units of 10$^{-15}$ Wcm$^{-2}$$\mu$m$^{-1}$ is shown in brackets after the
  date of observation.\label{fig3}}
\end{figure}

%figure 4
\begin{figure}
\plotone{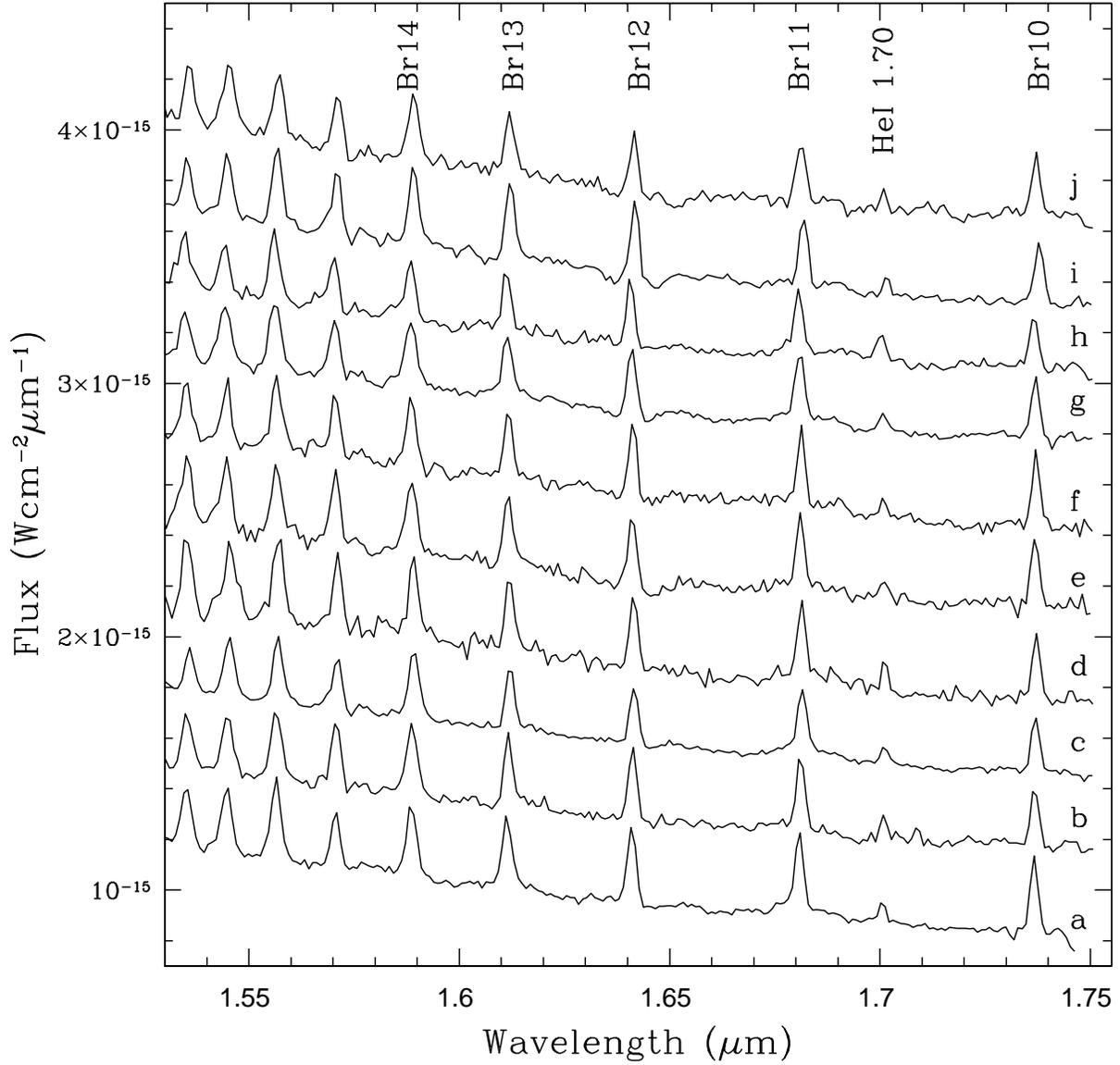}
\caption{Flux calibrated $H$ band spectra of X Per are displayed at different
  epochs with an offset between adjacent spectra for clarity. The amount of offset in
  units of 10$^{-15}$ Wcm$^{-2}$$\mu$m$^{-1}$ is shown in brackets after the
  date of observation, as shown below : (a) 2011 October 28, (b) 2011 October 30 (0.4), (c)
  2011 November 1 (0.8), (d) 2011 November 7 (0.95), (e) 2011 November 10
  (1.3), (f) 2011 November 21 (1.7), (g) 2011 December 1 (2.0), (h) 2011
  December 12 (2.35), (i) 2012 January 7 (2.5), (j) 2012 February 28 (2.8).\label{fig4} }
\end{figure}

%figure 5
\begin{figure}
\plotone{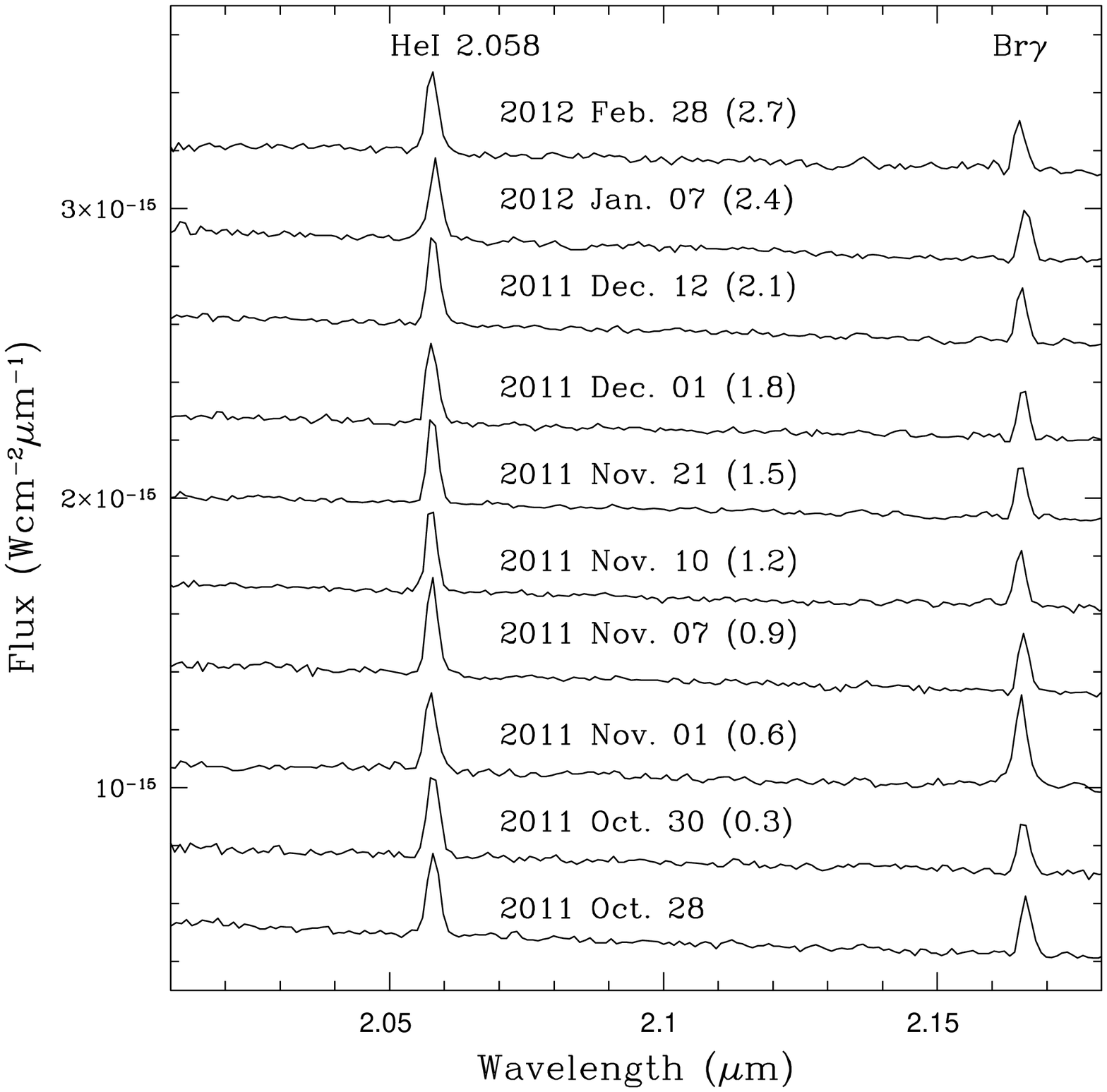}
\caption{Flux calibrated $K$ band spectra of X Per are displayed at different
  epochs with an offset between adjacent spectra for clarity. The amount of offset in
  units of 10$^{-15}$ Wcm$^{-2}$$\mu$m$^{-1}$ is shown in brackets after the
  date of observation.\label{fig5}}
\end{figure}

%figure 6
\begin{figure}
\plotone{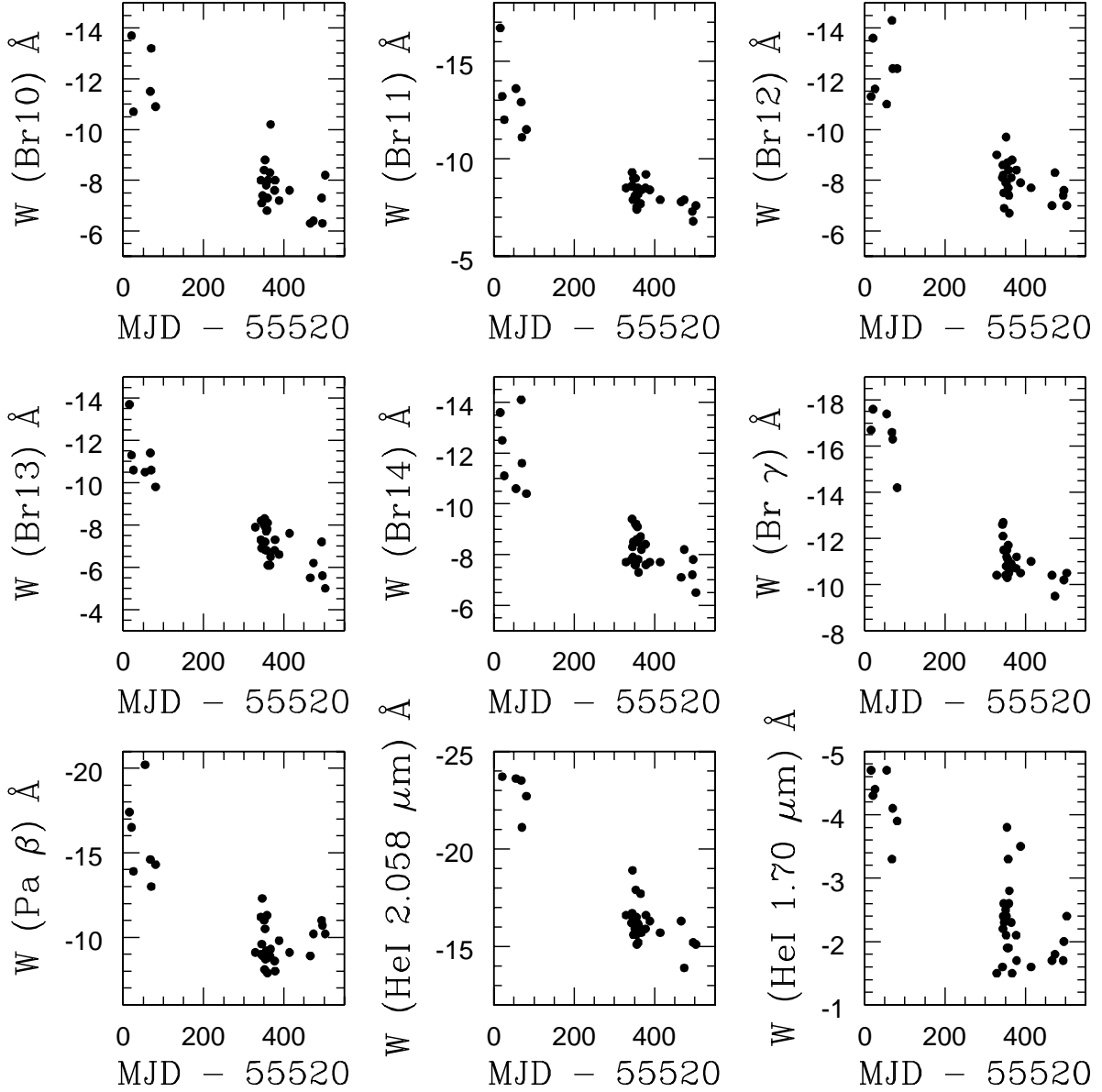}
\caption{The variation of equivalent width of major infrared emission lines
for various nights of observation is shown.\label{fig6}}
\end{figure}

%figure 7
\begin{figure}
\plotone{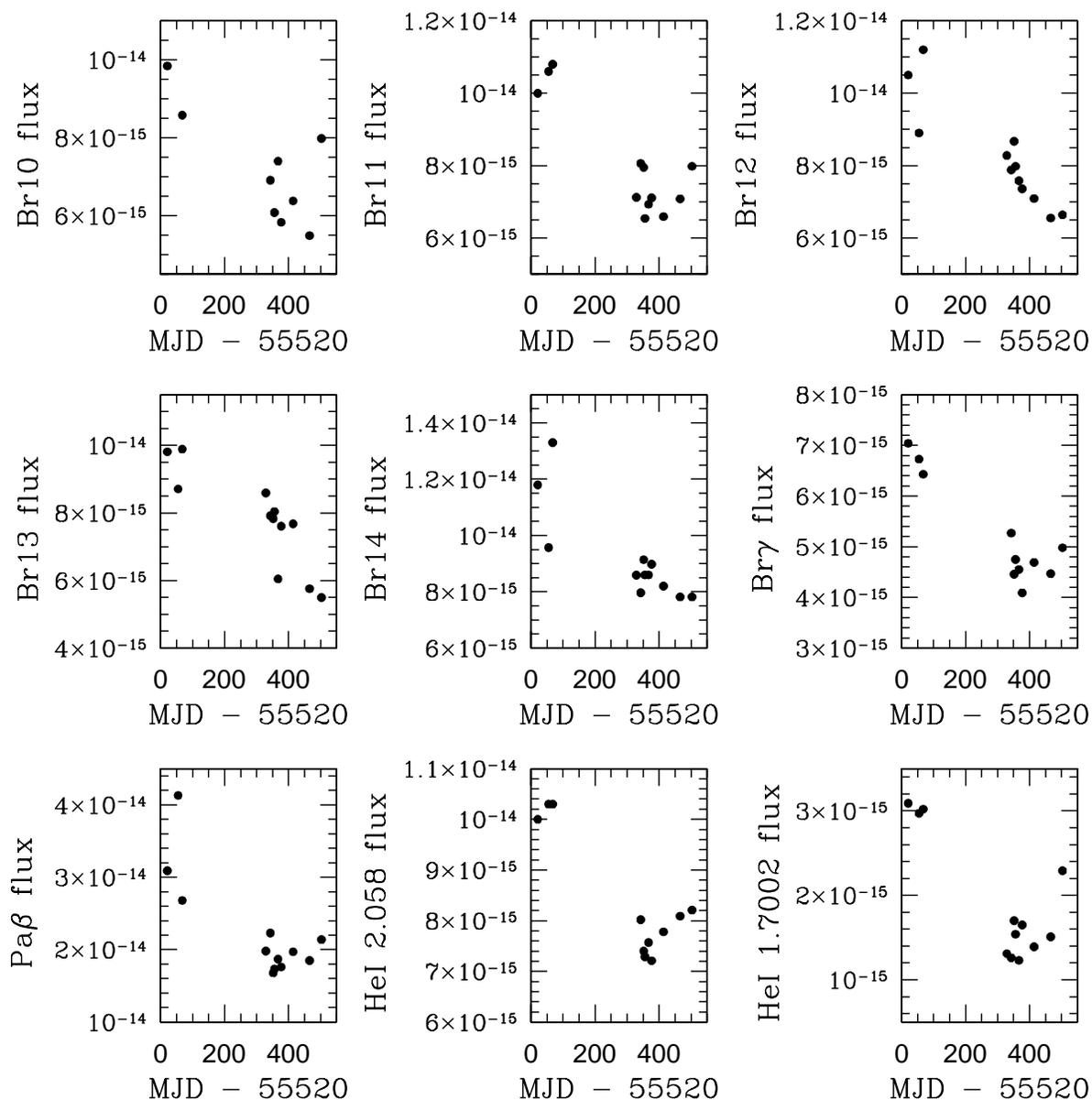}
\caption{The line flux variation of major infrared emission lines for various
nights is shown in units of Wcm$^{-2}$$\mu$m$^{-1}$.\label{fig7}}
\end{figure}

%figure 8
\begin{figure}
\includegraphics[bb= 35 1 297 297,width=5in,height=5.5in,clip]{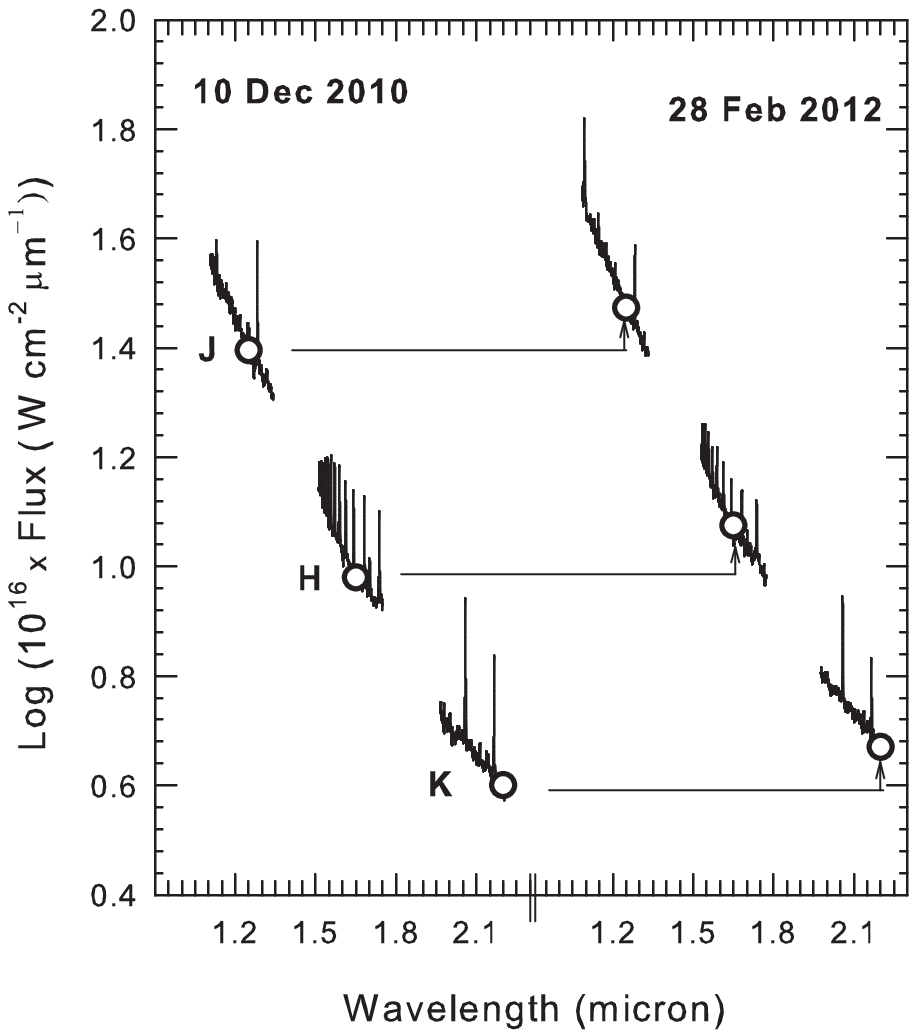}
\caption{Representative flux-calibrated spectra of 2010 December 10
and 2012 February 28 showing the general trend of the evolution of
the EW of the lines between 2010 and 2012. The spectra (continuous lines) are
flux calibrated using the corresponding broad band $JHK$ magnitudes
(shown in circles), dereddened using $E(B-V)$ = 0.39.
The arrowed lines between the spectra show the slight increase in
the continuum $JHK$ fluxes between 2010 and 2012,
whereas the spectra clearly show the anti-correlation in the
line strengths which are seen to decrease from 2010 to 2012.\label{fig8}}
\end{figure}

The spectra of X Per obtained earlier for 7 nights during 2010 December 05 -- 2011 February 8
(epoch 1) were analyzed and discussed in \citet{mat12b}.
We have combined equivalent widths of this epoch 1 data-set
with those of the present data to illustrate how the  equivalent width values evolve with time.
The equivalent width values obtained on 2011 January 28  are also
added in Table 2 as a sample representative of the previous set of observations.
The evolution of equivalent width values of Paschen $\beta$, Brackett $\gamma$, Brackett
10 -- 14, He{\sc i} 1.7002 $\mu$m, 2.058 $\mu$m emission lines are shown in
Figure 6. It can be seen that the EW values are reduced in the present set of
observations compared to the previous one.
In the present set of observations also a decreasing trend is visible which is
not as drastic in comparison with the previous data-set.
The trend is visible homogeneously in all the above mentioned emission lines.
What is clearly established in Figure 6 is that the equivalent
widths of the lines have systematically decreased with time
whereas the broad band continuum fluxes (shown in Figure 1) have
correspondingly increased during the same time.

We also consider how the line luminosities or line fluxes 
evolve with time ( we use the terms line luminosity and line flux
interchangeably). To determine the line fluxes we first 
flux calibrate the spectra with the corresponding  broad band magnitudes
which were measured simultaneously on the same night along with the spectra - 
these photometric magnitudes are listed in Table 1. Line fluxes were then extracted from these flux
calibrated spectra in a straight-forward way using IRAF  and these are shown
in Figure 7. As seen in the figure, the emission line flux values of all the
prominent lines decrease with time even as the continuum flux increases during
the same time. This decrease in the line fluxes is a genuine decrease and
shows that the anti-correlation of the equivalent widths with IR magnitudes
does not arise from a ``flux dilution'' effect. Flux dilution, also termed the ``veiling effect'', 
occurs whenever there is a source of continuum emission in addition to the
stellar photospheric emission and/or disk emission. The addition of a new
continuum source will dampen the flux of an emission or absorption line, 
decreasing its equivalent width without actually affecting the line flux. From
the mathematical definition of the equivalent width it can easily be seen that
by increasing the continuum flux F$_{c}$  by a factor of $N$, while keeping
the line flux F$_{line}$ constant, decreases the equivalent width $W$ by
1/$N$. Examples of the effect of ``flux dilution'' on photospheric absorption
lines may be seen for example in the papers by \citet{Aragona10} and \citet{Fremat06}.

For ease of comparison and visualization, of the evolution of the EW of the
lines between epochs 1 and 2, we have plotted in Figure 8 two representative
spectra of these respective epochs. These are for 2010 December 10
and 2012 February 28; the spectra are shown flux calibrated using the
corresponding broad band $JHK$ magnitudes dereddened using
$E(B-V)$ = 0.39 (these are shown as circles in Figure 8).
The arrowed lines between the spectra show the slight increase in
the continuum $JHK$ fluxes between 2010 to 2012,
whereas the spectra clearly show the anti-correlation in the line
strengths which are seen to decrease from 2010 to 2012.

To summarize, the continuum $JHK$ flux values are found to increase
(equivalently the J,H,K magnitudes are found to decrease) while the
equivalent width/line flux values are reduced with time in X per during
the period of observations. Hence it can be seen that an inverse correlation
exists  between the line equivalent widths (or line fluxes) and the continuum
emission. We examine possible physical mechanisms, some of which are suggested 
in the literature (for e.g. \citealt{sig13}),  that
could be responsible for producing such a trend.

\subsection{Discussion}

We have considered  three possible mechanisms to explain the observed
anti-correlated behavior of the  line and continuum strengths, viz., (i) the
recently proposed model by \citet{sig13} to account for such an
effect (ii) episodes of increased mass loss and (iii) the presence of a radiatively warped precessing disk.

\citet{sig13} have calculated model H$\alpha$ equivalent widths and 
UBV magnitudes for Be star disk models that grow in size and/or density with
time. They show that these simple models are consistent with the known Be star 
classes of positive and inverse correlations between long-term variations in
H$\alpha$ and V magnitude as identified by \citet{Harmanec83}. They show that the 
inverse correlation between H$\alpha$ and V magnitudes is expected at high
inclination angles due to the changing scale height of the growing disks. In their model, for Be stars observed at
an inclination angle $i$ less than $i$$_{crit}$, disk formation should result in a positive correlation,
while Be stars observed at $i$ greater than $i$$_{crit}$ should exhibit
an inverse correlation. The value of $i$$_{crit}$ is likely to depend on several parameters
and hence is difficult to assess accurately but a likely value is i$_{crit}$
$\sim$ 75$^{\circ}$. The authors however explain why this value may be  only
suggestive and should not be over-interpreted. In the case of X Per we likely
have a low inclination system with $i$ = 26$^{\circ}$ -- 33$^{\circ}$. This is
also the case with delta Sco ($i$ = 30.2$^{\circ}$; Meilland et al. 2011). 
So there may be some difficulty in reconciling the behavior of these  two
stars within the above model of \citet{sig13}. However, the
possibility exists that the infrared fluxes may be more sensitive to the
inverse correlation at lower inclinations since the vertical extent of the
disk could  be larger at IR wavelengths.

We next examine whether the observed anti-correlation might be
explained by episodes of increased mass loss. Such mass loss would
increase either the the electron density in the disk or the volume of
emitting gas - in either case the continuum and line  emission may be
expected to increase. This is because the  free-free (f-f) emission
from the disk will increase and in turn enhance the continuum;
independently and simultaneously  the strength of the lines arising
from the recombination process will also increase. This follows from
the  dependence of the continuum flux in the  f-f emission and the line flux
in the recombination process: in both processes the fluxes are
proportional to n$_e^{2}$V, where n$_{e}$ is the electron density and V is the
volume of the emitting gas \citep{ban01,sto95}. The possibility for the
strength of certain lines  to decrease - in this particular scenario - is
if the injection of matter increases the disk density to such levels that
these particular lines become optically thick. In which case the strength
of the lines may decrease - but a rigorous radiative transfer analysis must
be done to establish this which takes into account  the interplay between
optical depth at line-center ($\tau$) and the line strength.
From an observational point of view it is worth
pointing out that in several Be stars including X Per,  many of the  near-IR
lines are indeed optically thick in the Pa, Br, Pf and Hu series
\citep{mat12b,cla00,ste01,gra10} - especially the higher lines of the series.

In the case of delta Sco, \citet{mir03} noted a similar
anti-correlation between the visual and Balmer H$\alpha$ fluxes in the earlier
evolution of the delta Sco disk and suggested that it might be explained by episodes
of increased mass loss. Their reasoning was however based on a slightly
different approach than that offered above. From our point of view,
one of the arguments against the above mechanism is that the brightening
of the continuum,  by injection of matter into the disk,  should be seen
preferentially at the longer IR wavelengths, if caused by f-f flux increase,
rather than in the visual. This is expected because of the nature of the
dependence of the f-f emission with wavelength. However, in the present
case the light curves  in Figure 1 show a similar brightening of about
0.2 to 0.3 mag in each of the $V$, $J$, $H$ and $K$ bands. Such a variation could
arise from a geometric effect created by the formation or  removal of an
obscuration of the central star's light. A radiatively warped disk is
another possibility which could cause such a variation  as suggested by
\citet{car06} in the case of delta Sco.

\citet{por98} has shown how the radiation
field of the central star can warp the decretion disk of Be stars.
\citet{mar09}\footnote{http://stsci.academia.edu/RebeccaGMartin/Papers/1244536/}
also finds that  Be star disks can be  unstable to
radiation warping. If the density in the innermost parts of the disk is higher
than 10$^{-11}$ g cm$^{-3}$ then such parts of the disk are optically thick
and radiation driven warping can occur. Geometrically the warp  causes the
disk  to be effectively bent away from the original planar configuration
\citep{pri96,por97}. The effect of such a warp would place
material at higher latitudes and increase the area of the stellar disk
that could be blocked by the disk material along the line of sight.
This blockage, because of Thompson scattering of the incident continuum
radiation, would be expected to be more pronounced in low inclination
systems like X Per ($i$ = 26$^{\circ}$ -- 33$^{\circ}$) or
delta Sco ($i$ = 30.2$^{\circ}$; Meilland et al. 2011).
The continuum flux would hence decrease. However, since the warping causes
the disk surface  to curve out and thereby  offer a larger cross-sectional
area to the ionizing flux of the central star, more material  can be
photo-ionized which could lead to an increase in the  strengths of the lines.
Thus, by our reasoning, the current bright state is because the warped
portion of the disk is presently not in the line of sight (l.o.s). But if
this warped disk precesses, then the warped segment  should come into the
l.o.s in the future and a decline in the continuum should be seen then. It
needs to be seen whether this happens.

From an observational point of view, is there evidence for warped
precessing disks around Be stars? \citet{mar09} cites two possible
examples. First is the study of \citet{lar01} who
observed the Be/X-ray binary \object{A0535+26} and found a
beat frequency of the orbital period of the neutron star and the precession
period that could be due to a warped decretion disk around the Be star.
Second, \citet{neg01} observed the Be/X-ray transient
\object{4U0115+63/V635 Cas} in the optical, infrared and X-ray regimes and
found that the disk becomes unstable to, what they assumed to be radiative
warping, and then tilts and starts precessing.

To summarize, we have discussed the role of different mechanisms in
explaining the observed dependence of  the near-IR line strengths on 
the near-IR magnitudes in X Per. The observed anti-correlation is not easily
explained. We favor the presence of a radiation driven warped circumstellar 
disk to explain the observed behavior but it is possible that the underlying
mechanism originates from a different cause.

\acknowledgments

The research work at Physical Research Laboratory is funded by the Department
of Space, Government of India. We acknowledge with thanks the
variable star observations from the AAVSO
International Database contributed by observers worldwide and used in this
research. This work made use of $Swift/BAT$ data supplied by the UK Swift
Science Data Centre at the University of Leicester.

%table 2
\begin{deluxetable}{crrrrrrrrrrr}
\rotate
\tablecaption{Journal of the spectroscopic observations with the emission-line
equivalent widths shown in \AA.\label{tbl-2}}
\tablewidth{0pt}
\tablehead{
\colhead{Date} & \colhead{Exp. time (s)} & \colhead{Pa$\beta$} &
\colhead{Pa$\gamma$} & \colhead{Br10} & \colhead{He{\sc i}} &
\colhead{Br11} & \colhead{Br12} & \colhead{Br13} & \colhead{Br14} &
\colhead{Br$\gamma$} & \colhead{He{\sc i}}\\
\colhead{} & \colhead{$(J, H, K)$} & \colhead{} & \colhead{} & \colhead{} & \colhead{1.70 $\mu$m} &
\colhead{} & \colhead{} & \colhead{} & \colhead{} & \colhead{} &
\colhead{2.058 $\mu$m}
}
\startdata
2011 &  &   &  & & & & & & & & \\
Jan. 28\tablenotemark{a} & (40, 40, 40) & 12.8 & 17.0 &  13.1 & 4.1 & 10.8 & 12.1 & 10.2 & 10.8 & 16.2 & 20.6 \\
Oct. 14 & (60, 60, 60) & 9.1  & 15.3  & \nodata   & 1.5  & 8.5 & 9.0  & 7.9  & 7.7 & 10.4 & 16.6  \\
Oct. 28 & (60, 60, 60) & 11.2 & \nodata & 8.0 &  1.6 & 8.6 &  8.1 &  7.3 &  7.8&  12.6&  16.2  \\
Oct. 29 & (90, 90, 90) & 9.0  & 12.9  & \nodata  & 2.2 & 9.3 & 8.6  & 8.2  & 9.4 & 12.1 & 16.7  \\
Oct. 30 & (90, 90, 90) & 9.6  & \nodata  & 7.1  & 2.4  & 8.6  & 8.2  & 6.9  & 8.3 & 12.7 & 18.9  \\
Oct. 31 & (60, 60, 60) & 12.3 & 13.8 & \nodata  & 2.6 & 7.9 &  7.5 & \nodata  & 7.9&  11.5&  16.4  \\
Nov. 01 & (60, 60, 60) & 8.9  & \nodata  & 7.4  & 2.3 & 9.0  & 6.9  & 7.2  & 8.5 & 21.4 & 15.6  \\
Nov. 05 & (60, 60, 60) & 11.0 & \nodata  & 8.4 & 2.5 & 8.1 &  7.9 &  8.0 &  7.6&  10.4&  15.9  \\
Nov. 06 & (60, 60, 60) & 8.1  & \nodata  & \nodata & 2.1 & 9.0  & 9.7  & 8.3  & 9.2 & 10.8 & 16.1  \\
Nov. 07 & (60, 60, 60) & 10.5 & 15.3 &  8.8 & 2.4 & 7.9 &  7.9 &  7.2 &  7.6&  11.2&  17.9   \\
Nov. 08 & (60, 60, 60) & 8.7  & 14.8  &  \nodata  & 3.8 & 8.0 & 7.5  & 7.9  & 9.2 & 11.5 & 15.6  \\
Nov. 09 & (45, 50, 60) & 8.9  & 16.7  & \nodata  & 1.9 & 7.5 & 7.5  & 6.8  & 8.5 & 10.3 & 16.5  \\
Nov. 10 & (60, 60, 60) & 9.2  & 12.8  & 7.8  & 1.9 & 7.4  & 8.7  & 7.7  & 8.6 & 11.1 & 15.1  \\
Nov. 11 & (60, 60, 60) & 9.2  & 13.9  & \nodata  & 3.3 & 8.3  & 7.7  & 6.8  & 9.1 & 11.7 & 16.2  \\
Nov. 12 & (60, 60, 60) & 11.3 & 14.4 &  6.8 &  1.9 & 8.5 &  8.4 &  7.8 &  8.5&  10.8&  16.0  \\
Nov. 13 & (60, 60, 80) & 7.9  & 12.7  & 7.3  & 2.6 & 8.2  & 7.4  & 8.1  & 7.8 & 10.5 & 15.2  \\
Nov. 14 & (120, 60, 90) & 8.8  & 15.1  & 8.0  & 2.8 & 8.4  & 6.7  & 6.1  & 7.3 & 10.6 & 16.1  \\
Nov. 19 & (60, 90, 120) & 8.9  & 16.5  & 8.3  & 2.3 & 7.7  & 8.1  & 6.1  & 8.7 & 10.9 & 17.7  \\
Nov. 21 & (60, 60, 60) & 9.3  &  \nodata  & 10.2  & 1.5 & 8.4 &  8.8  & 6.5  & 8.2 & 10.8 & 15.7  \\
Dec. 01 & (60, 60, 60) & 8.6  & \nodata   & 7.6  & 2.1 & 8.5  & 8.4  & 6.8  & 8.4 & 10.7 & 15.9  \\
Dec. 02 & (60, 60, 60) & 8.0  & \nodata  & 8.0  & 1.7 & 9.2  & 8.4  & 7.3  & 7.6 & 11.2 & 16.6  \\
Dec. 12 & (40, 40, 60) & 9.8  & \nodata  & 7.2  & 3.5 & 8.4 & 7.9  & 6.6  & 7.7 & 10.5 & 16.3  \\
2012 &  &   &  & & & & & & & & \\
Jan. 07 & (60, 60, 60) & 9.1  & \nodata   & 7.6  & 1.6 & 7.9 & 7.7  & 7.6  & 7.7 & 11.0 & 15.7  \\
Feb. 28 & (60, 60, 60) & 8.9  & 15.2  & 6.3  & 1.7 & 7.8 & 7.0  & 5.5  & 7.1 & 10.4 & 16.3  \\
Mar. 07 & (60, 60, 60) & 10.2 & \nodata  & 6.4 & 1.8 & 7.9 &  8.3 &  6.2 &  8.2&  9.5&  13.9  \\
Mar. 27 & (60, 60, --) & 11.0 & 15.7 &  7.3 & 1.7 & 7.3 &  7.4 &  7.2 &  7.2& \nodata  & \nodata  \\
Mar. 29 & (60, 60, 60) & 10.7 & \nodata  & 6.3 &  2.0 & 6.8 &  7.6 &  5.6 &  7.8&  10.2&  15.2  \\
Apr. 05 & (60, 60, 60) & 10.2 & \nodata  & 8.2 & 2.4 & 7.6 &  7.0 &  5.0 &
6.5&  10.5&  15.1 \\
\enddata
\tablenotetext{a}{Data taken from \citet{mat12b}. See section 3.2 for more details.}
\end{deluxetable}

\end{document}